\documentclass[graybox]{svmult}

\usepackage{mathptmx}       % selects Times Roman as basic font
\usepackage{helvet}         % selects Helvetica as sans-serif font
\usepackage{courier}        % selects Courier as typewriter font
\usepackage{type1cm}        % activate if the above 3 fonts are
\usepackage{makeidx}         % allows index generation
\usepackage{graphicx}        % standard LaTeX graphics tool
\usepackage{multicol}        % used for the two-column index
\usepackage[bottom]{footmisc}% places footnotes at page bottom
\usepackage{hyperref}        %for hyperlinks
\usepackage{soul}            % for high-lighting of text
\usepackage{xcolor}
\hypersetup{colorlinks=true,urlcolor=blue}
\usepackage[square,numbers]{natbib}

\usepackage{amssymb}
\usepackage{xcolor}

\usepackage{caption}
\usepackage{subcaption}
\usepackage[colorinlistoftodos]{todonotes}

\usepackage{changes}  % Added for track-changes visualization, remove before submission

\usepackage{textcomp}
\usepackage{gensymb}

\usepackage[export]{adjustbox}

\setcitestyle{numbers}

\makeindex             % used for the subject index
\usepackage{gensymb}

\setcitestyle{authoryear,open={(},close={)}} 

\begin{document}

\title*{Ramaty High Energy Solar Spectroscopic Imager (RHESSI)}
\titlerunning{RHESSI}
\author{Brian Dennis\thanks{corresponding author}, Albert Y. Shih, Gordon J. Hurford, Pascal Saint-Hilaire}
\institute{Brian R. Dennis (\email{brian.r.dennis@nasa.gov}) and Albert Y. Shih (\email{albert.y.shih@nasa.gov}) \at Solar Physics Laboratory, Code 671, Goddard Space Flight Center, Greenbelt, Maryland 20771 \and Pascal Saint-Hilaire (\email{pascal@ssl.berkeley.edu}) and Gordon Hurford (\email{ghurford@berkeley.edu}) \at Space Sciences Laboratory, University of California Berkeley, 7 Gauss Way, Berkeley, California, 94720
}
\maketitle
\abstract{This chapter describes the X-ray and gamma-ray imaging spectroscopy capabilities of the Ramaty High Energy Solar Spectroscopic Imager (RHESSI). It also outlines RHESSI's major scientific accomplishments during the 16 years of operations from 2002 to 2018. These include unique contributions to solar flare research and to other aspects of solar physics (oblateness), astrophysics (magnetars), and Earth sciences (terrestrial gamma-ray flashes).
}


%\tableofcontents{}

\newpage

\section{Introduction}
\label{Introduction}
RHESSI \cite{2002SoPh..210....3L} combined both X-ray and gamma-ray imaging spectroscopy in a single instrument, with the goal of investigating particle acceleration and energy release in solar flares. It was launched on 5 February 2002 into a near-circular low-Earth orbit with an inclination of 38$\degree$. It operated almost continuously in a spin-stabilised Sun-pointing mode through April 2018, with brief off-pointing observations of the Crab Nebula and to measure the solar global emission from the quiet Sun \cite{2010ApJ...724..487HHannahHudsonHurfordApJ2010}. During its 16 years of operations, RHESSI detected more than 120,000 events that are documented in the RHESSI Flare List\footnote{\url{https://hesperia.gsfc.nasa.gov/hessidata/dbase/hessi_flare_list.txt}}. Over 23,000 of them have
detectable emission above 12 keV, 979 above 50 keV, 468 above 100 keV, and 223 above 300 keV; 27 events show gamma-ray line emission. The RHESSI observation of the GOES X-class flare, SOL2002-07-23T00:35, yielded the first-ever spectrally resolved measurements of solar gamma-ray lines and the first-ever gamma-ray image of a solar flare. See \cite{2009ApJ...698L.152SShihLinSmithApJL2009}
for a summary of all gamma-ray events seen with RHESSI as of the Cycle 23/Cycle 24 solar minimum. 

There were five successful anneals of the germanium detectors, the first in November 2007 and the last in April 2016. These were month-long procedures that involved heating the detectors to close to 100~$\degree$C for a period of close to a week. They were necessary in order to restore the sensitive volume and energy resolution that become degraded by the accumulation of radiation damage.

RHESSI covered energies from as low as 3 keV to as high as 17 MeV with a FWHM resolution of $\sim$1 keV at the lowest energies increasing to $\sim$5 keV at 5 MeV. Imaging with a full-Sun field of view was possible with the finest angular resolution of $\sim$2 arcsec up to 100 keV, 7 arcsec to 400 keV, and 36 arcsec in the 2.22 MeV neutron capture line. A temporal resolution as fine as tens of ms was possible for count-rate variations but was generally limited for imaging to $\geq$4 s by the 15 rpm spacecraft rotation rate and the counting statistics for any given flare.

\section{Objectives}
\label{Objectives}

{\em RHESSI} \cite{2002SoPh..210....3L} was designed with the primary scientific objective of understanding particle acceleration (both electrons and ions) and explosive energy release in magnetised plasmas at the Sun. This was to be achieved through both X-ray and gamma-ray imaging spectroscopy. Observations of the bremsstrahlung hard X-rays reveal the location and energy spectra of the flare-accelerated electrons, and the nuclear gamma-rays provide information on the accelerated protons and heavier ions.

Previous X-ray measurements, coupled with measurements at other wavelengths, had shown that in order to achieve the goal for electron acceleration, hard X-ray imaging was required with an angular resolution in the few-arc-second range at energies from below 10 keV to several hundred keV. Such an angular resolution was necessary to separate chromospheric footpoint sources and distinguish them from the hot coronal sources. It was also necessary to determine the spectra of the bremsstrahlung emission from non-thermal electrons at energies well above that of the thermal bremsstrahlung emission from plasma with temperatures ranging from below 10 MK to in excess of 50 MK. In addition, it was necessary to have an effective sensitive area of $\sim$100 cm$^2$ in order to detect weaker events while also avoiding saturation for the largest events. This corresponds to a requirement for handling a dynamic range of some $10^5$ in flare X-ray intensity from small GOES A-class events to the largest X10 events. 

In order to achieve the goal of understanding ion acceleration, gamma-ray imaging and spectroscopy was required at energies between hundreds of keV and tens of MeV. At these energies, lines and continuum emission are produced through nuclear reactions of accelerated protons and heavier ions interacting with the ambient thermal ions in the solar atmosphere. Large volume detectors were required with sufficient sensitivity and fine energy resolution in this difficult energy range where the photon penetrating power becomes so large.

\begin{figure}[pht]
\includegraphics[width=0.8\textwidth]{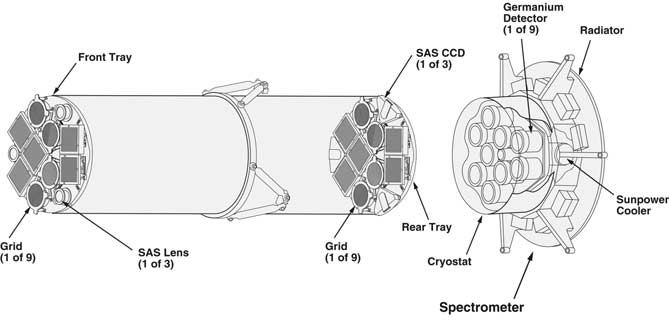}
\centering \caption{\footnotesize Perspective of {\em RHESSI} showing the components necessary for imaging. Two essentially identical sets of nine grids are mounted on front and rear grid trays, with all the grids made of tungsten except for the finest grid pair that was made of molybdenum for ease of etching. A corresponding set of nine cooled germanium detectors is mounted behind the rear grids. The solar aspect system (SAS) consists of three lenses mounted on the front grid tray that focus optical images of the Sun onto three linear diode arrays (labeled SAS CCD in the figure) on the rear grid tray. It provides sub-arcsec knowledge of the radial pointing with respect to Sun centre. Two optical roll angle systems (RASs, not shown) pointed perpendicular to the spin axis detect multiple stars each spacecraft rotation and provide the necessary roll angle information. The combined SAS and RAS data enables the absolute orientation of the grids to be determined on millisecond time scales and allows X-ray sources to be located on the solar disk to sub-arcsec accuracy \cite{Fivian2002}. From \cite{Hurford2002}.}
\label{RHESSI Perspective}
\end{figure}

\begin{figure}[pht]
\includegraphics[angle=90, width=0.8\textwidth]{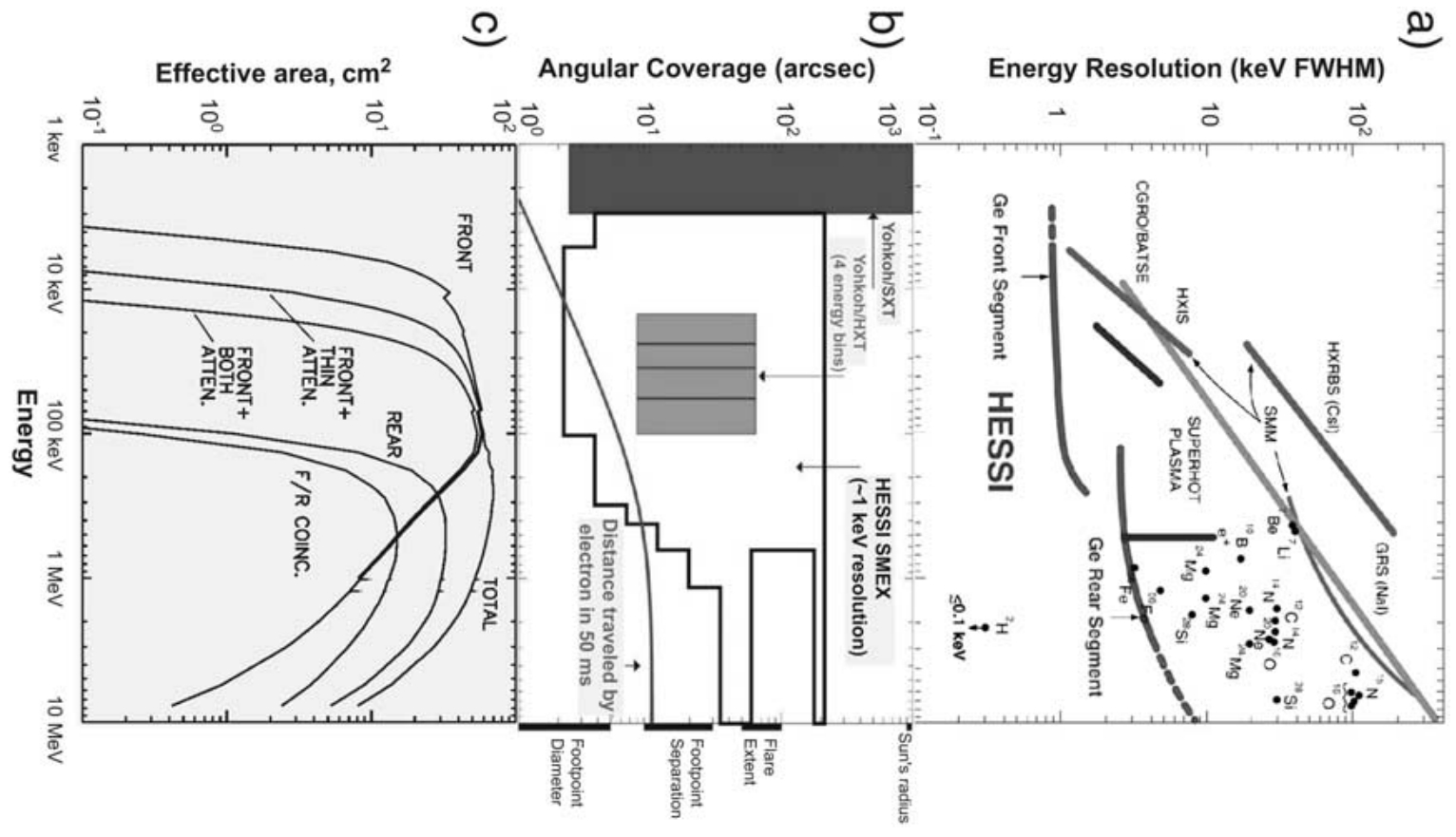}
\centering 
\caption{\footnotesize
 Summary of RHESSI's spectral and angular resolution and its effective area as functions of the incident photon energy. (a) RHESSI’s full width at half maximum (FWHM) energy resolution (lower curves), compared with that of previous instruments (upper lines). and the expected widths of the predicted gamma-ray lines. (b) RHESSI’s angular source size coverage versus energy, compared to Yohkoh's Soft X-ray Telescope (SXT) and Hard X-ray Telescope (HXT). (c)~RHESSI’s effective (photopeak) area as a function of energy for solar X-rays and gamma-rays. The three curves below 100 keV show the effect on the front segment response (summed over all nine detectors) of no attenuators, thin attenuators, or both thick and thin attenuators inserted over the detectors. The curves at higher energies show the effective area for the front and rear segments separately, and for photons that leave energy in both the front and rear segments of the detectors (i.e., that produce a front/rear coincidence).
From \cite{2002SoPh..210....3L}.}
\label{RHESSI Capabilities}
\end{figure}

After reviewing the different available techniques for imaging X-rays and gamma-rays (see the chapter in this volume on Grid-based Imaging of X-rays and Gamma-rays with High Angular Resolution), it was realised that the only practical method of satisfying these observational requirements within the constraints of NASA's Small Explorer (SMEX) program was to use collimator-based Fourier-transform imaging \cite{Hurford2002}. With only a single instrument involved, it was possible to simply rotate the whole spacecraft in a spin-stabilised configuration to achieved the desired temporal modulation of the incident flux. The resulting RHESSI instrument design is shown in Fig.~\ref{RHESSI Perspective} with its capabilities summarised in Fig.~\ref{RHESSI Capabilities}.

\section{Design and Capabilities}
\label{Design}

\subsection{Spectroscopy}

\label{Spectroscopy}
RHESSI covered the broad energy range from $\sim$3 keV to 17 MeV with nine ultra-pure coaxial germanium detectors cooled to $\sim$70~K \cite{Smith2002}. The fine ($\sim$1~keV~FWHM) X-ray energy resolution of these detectors shown in Fig.~\ref{RHESSI Capabilities}(a) has allowed for the detection and characterisation of the iron-line complex at $\sim$6.7 keV and the exponentially falling thermal bremsstrahlung spectra and power-law non-thermal spectra that can be as flat as $\epsilon^{-2}$ (where $\epsilon$ is the X-ray photon energy) or as steep as $\epsilon^{-10}$. The $\sim$2 to 5 keV gamma-ray energy resolution allows all of the nuclear lines, except for the intrinsically narrow neutron capture line at 2.223 MeV, to be fully resolved and line shapes to be determined.

A single cylindrical large-volume (7.1 cm diameter x 8.5 cm long) germanium detector \cite{Smith2002} was used behind each collimator. These detectors were able to cover the full energy range required to satisfy both the electron and ion acceleration goals. This was achieved by electronically separating each detector into two segments, a $\sim$1~cm thick front segment, optimised for lower energy photons at energies below a few hundred keV, and a $\sim$7~cm thick rear segment, optimised for higher energy photons up to $>$10~MeV.

\subsubsection{Dynamic Range}
\label{Dynamic Range}

RHESSI operated with minimal detector saturation and pulse pile-up over a wide dynamic range in flux level. This allowed coverage ranging from GOES A-class events and up to the most powerful GOES X10 flares. This was achieved in two ways. Firstly, the electronic segmentation of each germanium detector meant that in the most intense flares, the rear segments were shielded from the intense soft X-ray fluxes and could still be used for hard X-ray and gamma-ray imaging spectroscopy. Secondly, one or two thin aluminium disks were automatically moved above each detector to absorb an energy-dependent fraction of the incident X-ray flux when the front segment count rate exceeded a threshold value of $\sim$10$^4$ counts s$^{-1}$ detector$^{-1}$. This reduced the count rate in the front segments to an acceptable level but still allowed coverage to energies down to 6 keV. 

The effective sensitive area is shown in Fig.~\ref{RHESSI Capabilities}(c) as a function of energy for the front and rear detector segments separately and summed together, and for different attenuator states. It rises from a threshold at 3 keV with all attenuators removed or 6 keV with both attenuators in place. It reaches a peak of $\sim$60 cm$^2$ at 100 keV and falls to $\sim$10 cm$^2$ at 10 MeV.

\subsection{Imaging}
\label{Imaging}
The imaging capability was achieved with a bi-grid tungsten collimator\footnote{The collimator with the finest angular resolution was made of molybdenum because of etching limitations} over each of the nine germanium detectors to modulate the incident photon flux as the spacecraft rotated at $\sim$15 rpm. This provided the temporal information needed for the Fourier-transform technique that is used to reconstruct the X-ray and gamma-ray images \cite{Hurford2002}.

The pitch of the slits in the grids of the nine sub-collimators ranged from as fine as 34 microns to 2.75 mm increasing in steps of $\sqrt3\times pitch$. The thickness of the grids ranged from 1.2 mm for the finest grids to 3 cm for the coarsest grid. The range of source sizes that could be imaged with the modulated count rates in the detectors behind each of these sub-collimators is shown in Fig.~\ref{RHESSI Capabilities}(b). It extends from $\sim$2~arcsec (FWHM) as determined by the finest grids up 180 arcsec as determined by the coarsest grids. The maximum energy at which imaging with a given resolution is possible is determined by the energy at which the corresponding grid becomes transparent to photons of that energy. The smallest source dimension that can be imaged is thus 2 arcsec up to $\sim$100 keV increasing to $\sim$20~arcsec at 1 MeV. Thicker tungsten grids for two of the coarser collimators with angular resolutions of 35 and 183 arcsec allowed for modulation at the highest energies and enabled RHESSI to make the first ever gamma-ray images in the neutron capture line at 2.223~MeV \cite{2003ApJ...595L..77H,2006ApJ...644L..93H} (see Section \ref{GammaRayOverlay}).

The field of view through all the individual grids was $\geq1\degree$ so that a flare could be imaged no matter where on the visible disk it occurred.

\subsubsection{Rotating Modulation Collimators}
\label{RMCs}

At X-ray energies up to $\sim$1~MeV, the use of rotating modulation collimators (RMCs) allowed for the measurement of over a hundred spatial frequency (i.e., Fourier) components, covering a broad range of angular size scales from 2 to 180 arcsec. The only other solar X-ray imager using RMCs was the Solar X-ray Telescope (SXT) on the Japanese Hinotori spacecraft, launched in 1981 \cite{Enome1982}. It consisted of two orthogonal bi-grid modulation collimators on a spacecraft rotating at 4.3 rpm. The Hard X-ray Telescope (HXT) on Yohkoh also used bi-grid modulation collimators but on a 3-axis stabilised spacecraft. It measured 32 Fourier components covering spatial scales from 8 to 20 arcsec \cite{Kosugi1991}.

Another advantage of using bi-grid rotating modulation collimators is that the detectors need no spatial resolution -- all the imaging information is encoded in the temporal modulation of the detector counting rates as the spacecraft rotates. Thus the same detectors used for spectroscopy as described in Section \ref{Spectroscopy} could be used to measure the temporally modulated count rates generated by each of the nine modulation collimators.

\subsubsection{Imaging Concept}
\label{Imaging Concept}

\begin{figure}[pht]
\includegraphics[width=1.0\textwidth]{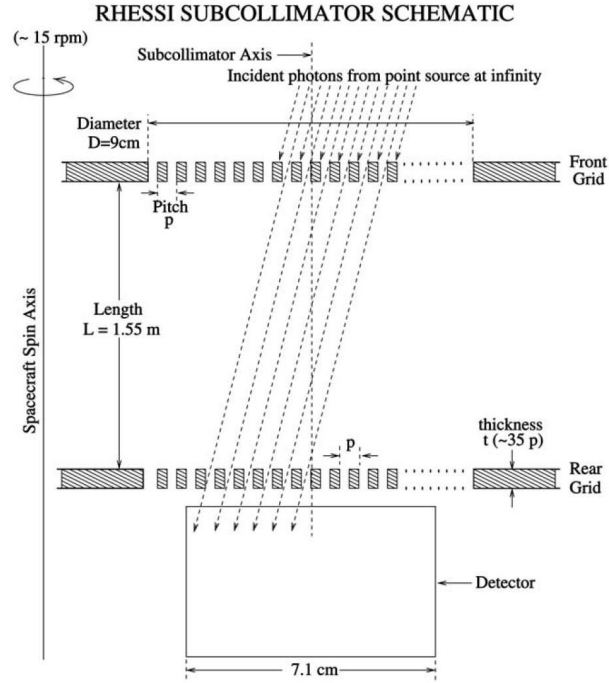}
{\vskip-1.5in}
\centering \caption{\footnotesize Geometry of the {\em RHESSI} imaging process. Light from a distant point source passes through two identical grids, each with a slit/slat pitch of $p$ and separated by a distance $L=1.55$~m. A~detector records a photon only if it passes through both grids; thus the detected flux depends on the orientation of the grids relative to the source direction.  In the spacecraft frame, this direction changes continually due to spacecraft rotation, thus providing a temporally modulated signal that provides information on the source direction. From \cite{Hurford2002}.}
\label{RHESSI Geometry}
\end{figure}

The basic {\em RHESSI} imaging concept is described in \cite{Hurford2002}. The {\em RHESSI} instrument encoded imaging information through the RMC technique illustrated in Fig.~\ref{RHESSI Geometry}. A set of parallel occulting grids rotates with the spacecraft, sweeping possible observing directions across the solar disk and rapidly modulating the detected photon flux in the process. Imaging information is encoded in the amplitudes and phases of the X-ray counting rate modulations in the different detectors as the spacecraft rotates at nominally 15~rpm. {\em RHESSI} is thus termed a ``Fourier imager'' because it provided imaging information through measurement of the spatial (or, more accurately, angular) Fourier components of the X-ray source(s) that produce these modulated time profiles.

\subsubsection{Image Reconstruction}
\label{Imaging Reconstruction}
Various computational techniques have been devised to generate images of the X-ray sources from telemetered data. These are discussed in detail in \cite{Piana2022}. They include methods based directly on the measured counting rates in the nine germanium detectors and those based on the visibilities derived from these counting rates. The former methods include CLEAN, forward fitting, and Pixon; the latter include forward fitting, Bayesian optimisation, Maximum Entropy, etc. One additional capability is to directly generate maps of the flux of the energetic electrons that produced the measured bremsstrahlung X-rays \cite{2007ApJ...665..846P}.

\subsubsection{{\em RHESSI} Imaging Example}
\label{RHESSI Imaging Example}

\begin{figure}[pht]
\begin{center}
\includegraphics[width=0.30\textwidth, center, 
    angle = 0, trim = 200 0 200 0]
    {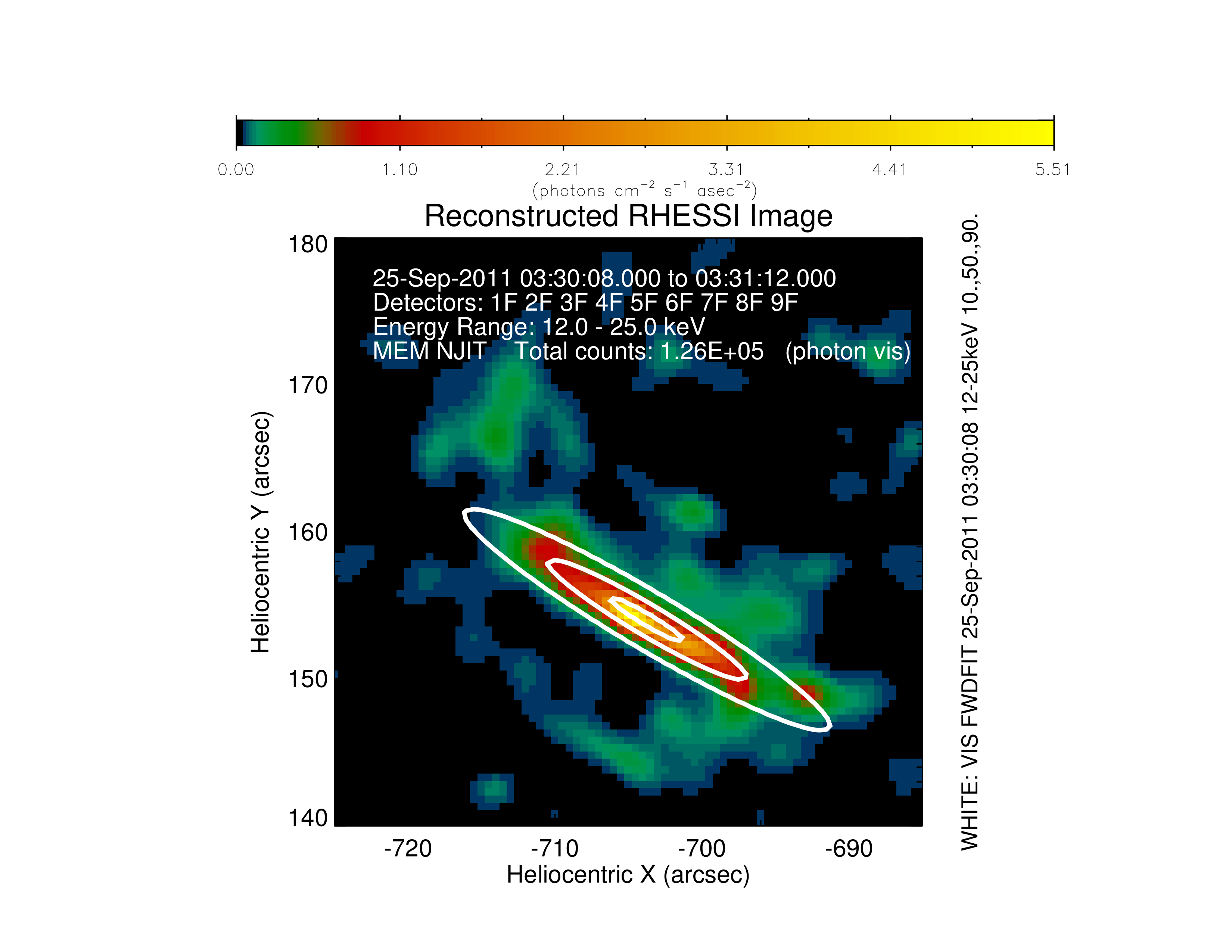}
\includegraphics[width=0.50\textwidth, center,
    angle = 0, trim = 0 0 0 0]
{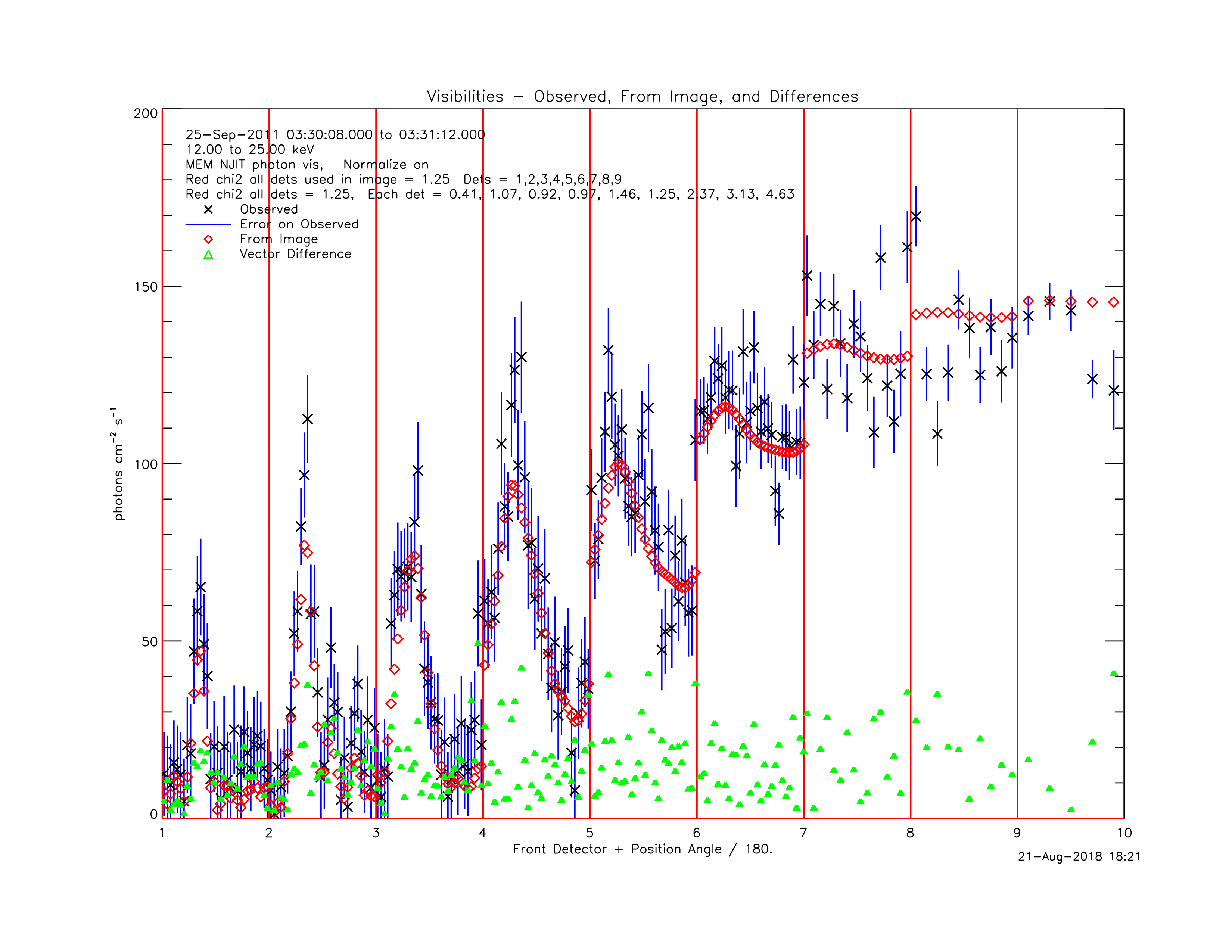}
\includegraphics[width=0.5\textwidth,center, 
    angle = 0, trim = 0 50 0 0]
    {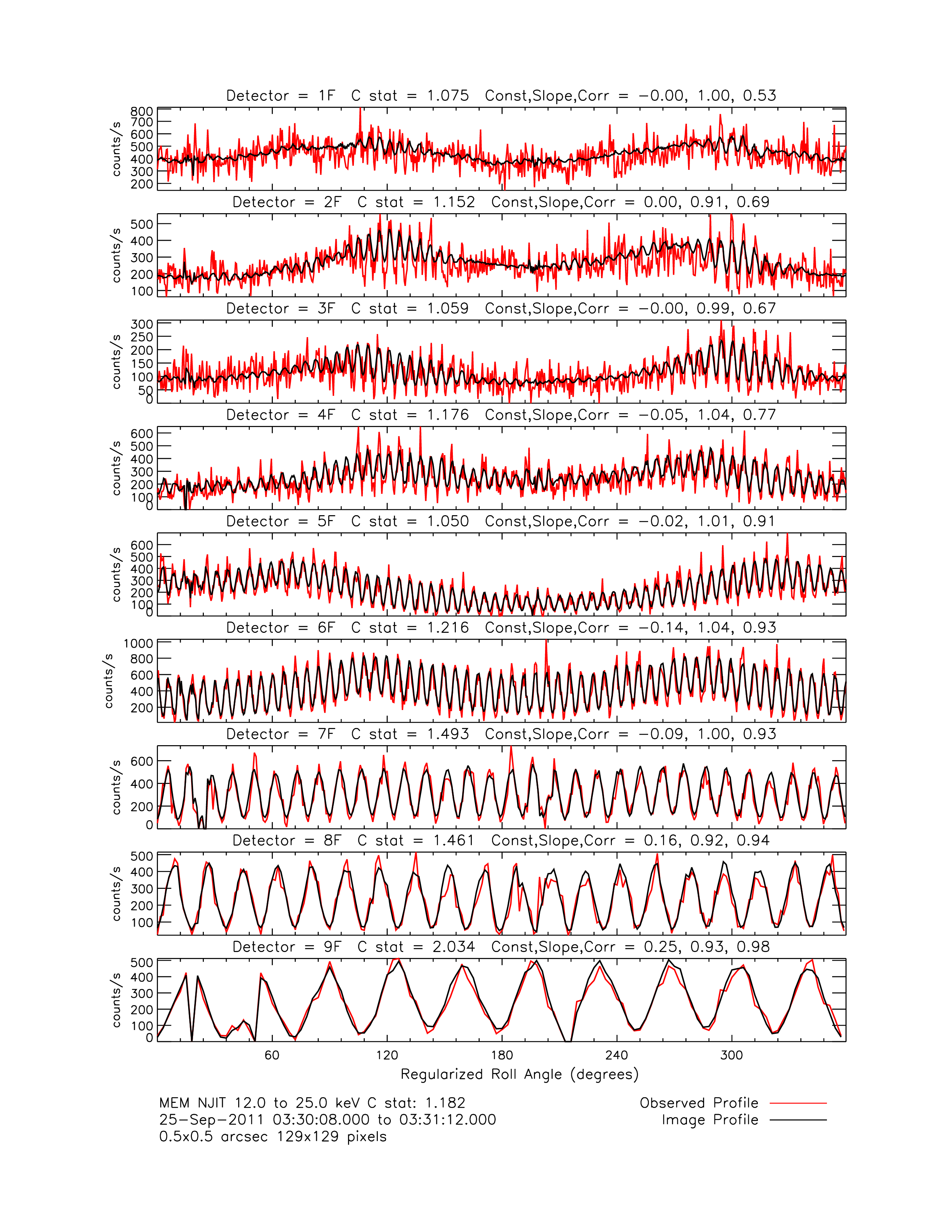}
\end{center}
\caption{{\it Top}: {\em RHESSI} colour image in the 12--25~keV energy range made using the MEM\_NJIT reconstruction method. White contours show the extent of a best-fit elliptical Gaussian source determined with the VIS\_FWDFIT procedure. {\it Middle}: Visibility amplitudes plotted against position angle for each of the nine {\em RHESSI} detectors, shown as black crosses with $\pm1\sigma$ error bars in blue. The amplitudes determined from the image reconstructed using MEM\_NJIT are shown as red diamonds, with the vector difference amplitudes shown as green triangles. {\it Bottom}: Count rates (in red) in the front segment of each of the nine germanium detectors for the same time interval and energy band as for the image. These are plotted versus ``regularised roll angle,'' defined as the spacecraft roll angle corrected for the offset between the spin axis and the mean sub-collimator optical axis. Overlaid in black are the expected count rates predicted from the image shown to the left. From \cite{Dennis2019}.}
\label{RHESSI Imaging Example Figures}
\end{figure}

An example illustrating {\em RHESSI}'s capability to image a simple asymmetric source with different extents in perpendicular directions was presented in \cite{Dennis2019} and is shown in Fig.~\ref{RHESSI Imaging Example Figures}. The 12--25~keV image shown in the top left panel was made during the GOES C7.9 class flare that occurred on 25 September 2011 in NOAA active region 11302 at N12E50, peaking at 03:32~UT. The white contours show the image made with the VIS\_FWDFIT image reconstruction method (see \citet{Piana2022} for detailed descriptions of this and other methods for reconstructing images from the RHESSI observations) used under the assumption that the source is a single elliptical Gaussian. The best fit to the visibilities $V(u,v)$ was obtained with FWHM dimensions of $15.8\pm0.4$~arcsec in length and $2.0 \pm 0.2$~arcsec in width. The counting rates and visibility amplitudes for each of the nine germanium detectors, shown in Fig.~\ref{RHESSI Imaging Example Figures}, illustrate how changes in the modulations are produced by this relatively simple source and how inspection and analysis of these modulations leads to the reconstructed image. The bottom panel in  Fig.~\ref{RHESSI Imaging Example Figures} shows the count rates for each detector separately plotted as a function of ``regularised roll angle,'' defined as the spacecraft roll angle corrected for the offset between the spin axis and the mean sub-collimator optical axis. To build up statistics, the count rates are ``stacked,'' i.e., summed modulo the spacecraft spin period ($\sim$4~s) for the 64~s interval duration used for the image.

Significant sinusoidal modulation is evident in the plots for all detectors. The detectors behind the sub-collimators with the coarser grids ($\#5$~to~$\#9$) show modulation at all roll angles, while the detectors behind the finer grids ($\#1$~to~$\#4$) show modulation over two limited ranges in regularised roll angle ($\sim$90$^\circ-150^\circ$ and $\sim$280$^\circ-300^\circ$). This is the expected modulation signal for an asymmetrical source with a width much smaller than its length, oriented as shown in the top left panel. The agreement between the measured and predicted rates is an indication of how well the reconstructed image matches the data. This agreement is quantified with the Cash or C-statistic \cite{Cash1979}, given as a separate value for each detector and as an overall value for all detectors together.

Another way of displaying the modulation in the different detectors is shown in the top right panel of Fig.~\ref{RHESSI Imaging Example Figures}, sometimes called a ``Hurford'' plot after its originator. Here, the amplitudes of the visibilities for each detector are plotted as a function of position angle, defined as the spatial direction of each grid response referenced to solar North. This is essentially the same as plotting the amplitudes of the count-rate oscillations seen in the top right panel, except that the data are plotted for only a half rotation from $0^\circ$ to $180^\circ$ (with the second half rotation assumed to be identical and hence added to the first, an option known as ``combine conjugates''). Here again, the characteristics of an asymmetric source are dramatically evident, with the peaks in the visibility amplitudes for each detector showing the position angle of the smallest dimension of the source and the valleys showing the position angle of the largest dimension. Since the peak is evident even in detector~\#1, we know that the smallest source dimension must be close to the 2~arcsec resolution of the finest grids. The count rates in the valleys are similar in detectors~\#1, 2, and~3, with detector~\#5 having $\sim$50\% of the modulation amplitude of detectors~\#8 and~9. This indicates that the longest source dimension must be similar to the 20~arcsec resolution of detector~\#5, i.e., similar to the value found from VIS\_FWDFIT. From this analysis, another quantitative indication of how well the image fits the count rates is given by the reduced $\chi^2$ values computed from the measured and predicted visibility vectors weighted by the statistical uncertainties. These values are listed in the figure for each detector and for all detectors together.

\section{Scientific Legacy}
\label{Legacy}

A discussion of the major scientific accomplishments of the RHESSI mission is given in the last Senior Review proposal submitted to NASA in 2017\footnote{\url{https://hesperia.gsfc.nasa.gov/rhessi3/docs/senior_review/2017/senior_review_proposal_2017.pdf}}. The results were broken down into the following four major 
elements plus two additional components derived from RHESSI’s serendipitous optical and non-solar capabilities:
\begin{enumerate}
\item Evolution of solar eruptive events (SEEs) 
\item Acceleration of electrons 
\item Acceleration of ions
\item Origin of thermal plasma
\item Optical Sun
\item X-ray and gamma-ray sources of both terrestrial and astrophysical origin.
\end{enumerate}

A comprehensive review of RHESSI's scientific achievements can be found in the Springer monograph entitled High-Energy Aspects of Solar Flares\footnote{\url{https://link.springer.com/book/9781461430728}} and also published in Space Science Reviews \cite{2011SSRv..159....3DDennisEmslieHudsonSSRv2011}\footnote{The flares that are included in this review are discussed in RHESSI Science Nugget \#143 accessible from  \url{https://hesperia.gsfc.nasa.gov/rhessi3/news-and-resources/results/nuggets/}}.
We present here a top-ten list\footnote{This list was presented in the RHESSI's Tenth Anniversary Science Nugget \#169 accessible from
\url{https://hesperia.gsfc.nasa.gov/rhessi3/news-and-resources/results/nuggets/}} of the iconic results that show RHESSI's unique contributions to solar flare research. Three additional items are added that resulted from RHESSI's serendipitous capabilities in other aspects of solar physics (oblateness), in astrophysics (magnetars), and in Earth sciences (terrestrial gamma-ray flashes). 

%\subsection{Solar Flare Results}

\subsection{Discovery of Gamma-Ray Footpoint Structures}

\begin{figure}[pht]
\begin{center}
\includegraphics[width=0.8\textwidth, 
    angle = 0, trim = 0 0 0 0]
    {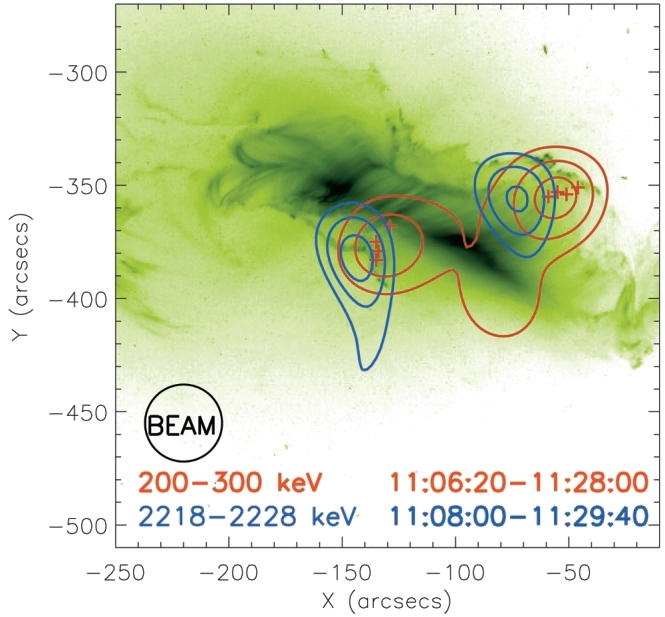}
\end{center}
\caption{Overlay of gamma-ray and X-ray contours on a TRACE 195 \r{A} image of the 2003 October 28 flare. From \cite{2006ApJ...644L..93H}.}
\label{GammaRayOverlay}
\end{figure}

Prior to RHESSI, very little was known about where and how energetic ions are accelerated in solar flares. This is important since these ions may contain as much energy as energetic electrons, both carrying of order 10--50\% of the total energy released in a flare. RHESSI has provided the first-ever imaging of energetic flare ions through their nuclear gamma-ray line emission \cite{2003ApJ...595L..77H,2006ApJ...644L..93H}. In the largest flare, shown in Fig.~\ref{GammaRayOverlay}, double ion footpoints seen in gamma-rays are detected straddling the loop arcade, closely paralleling the electron footpoints seen in hard X-rays.
These observations show that the energetic ions are located in small footpoint sources in the vicinity of the flare. This is inconsistent with acceleration over a broad region by a shock front propagating far from the flare as suggested for solar energetic particle (SEP) acceleration by shocks driven by fast CMEs. This strongly suggests that flare ion acceleration is similar to flare electron acceleration, with both possibly related to the process of magnetic reconnection.  As shown in the figure, however, the ion footpoints are displaced from the electron footpoints by $\sim$10--20 Mm, for reasons unknown.  

RHESSI has also provided the first high-resolution spectroscopy of flare gamma-ray lines \cite{Smith2003}. 
Detailed analysis revealed mass-dependent Doppler red shifts of order 1\%, indicating that the emitting ions were traveling downward at an angle of $\sim$30$\degree$ to $40\degree$ to the vertical -- likely along tilted magnetic field lines. 
This result, combined with the detection of gamma-ray footpoint emission, shows clearly that the ions must have been accelerated over a relatively small volume in the corona, on closed field lines in the primary flare energy-release region. 

\subsection{Energy Content and Spectrum of Flare Energetic Electrons}

A crucial question for flares is, how much of the energy released goes into particle acceleration? For the first time, RHESSI was able to show unambiguously that the power-law spectrum of energetic electrons extends down to $\sim$20 keV for many flares, and therefore these electrons must contain a large fraction, $\sim$10--50\%, of the total energy released in those flares.  RHESSI’s uniquely fine energy resolution ($\sim$1~keV FWHM) was sufficient to resolve the steep high-energy fall-off of the hot flare thermal continuum, allowing for the accurate determination of the energy above which the hard X-ray emission must be non-thermal (e.g., \cite{2003ApJ...586..606H}).

\begin{figure}[pht]
\begin{center}
\includegraphics[width=0.8\textwidth, 
    angle = 0, trim = 0 0 0 0]
    {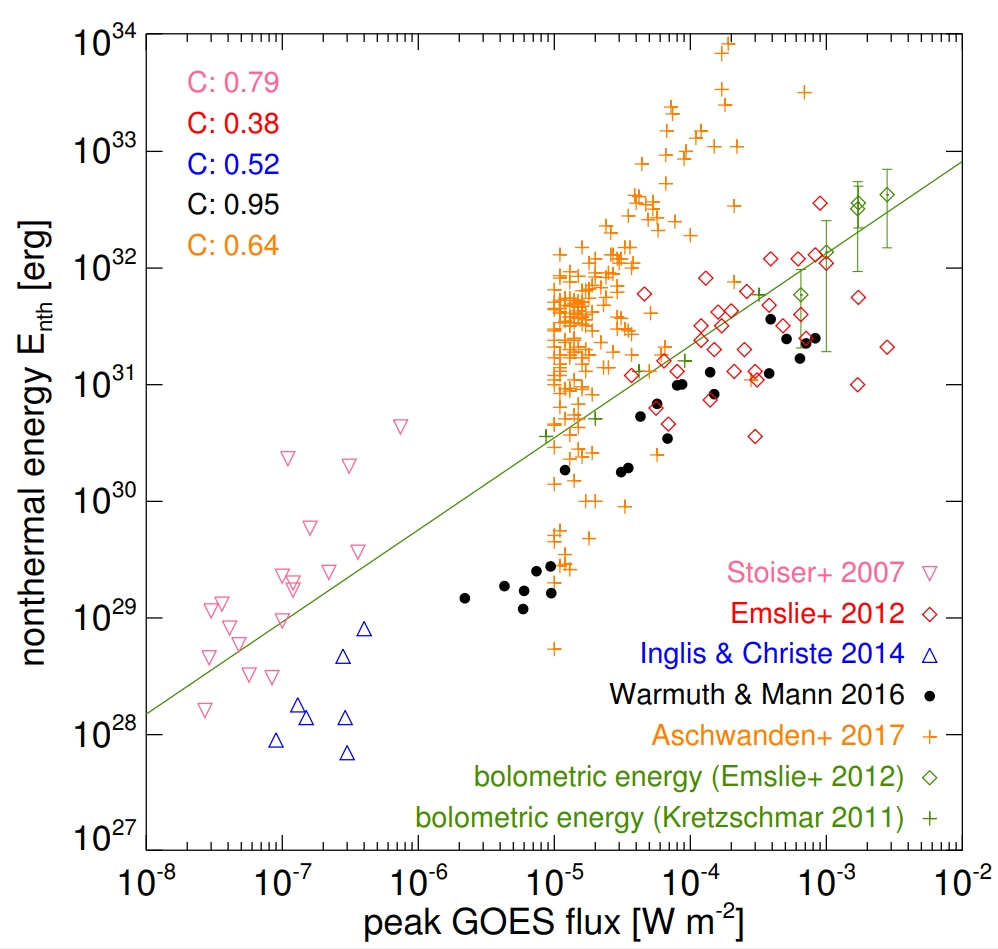}
\end{center}
\caption{Energy in non-thermal electrons,
E$_{nth}$, plotted versus peak GOES flux for all five studies.  For comparison, the total radiated energies E$_{bol}$ derived from SORCE/TIM (green diamonds) and SOHO/VIRGO (green crosses). The
green line is a power-law fit to these data points). From \cite{2020A&A...644A.172WWarmuthMannA&A2020} with permission of A. Warmuth.}
\label{WarmuthMann}
\end{figure}

For bright solar flares, RHESSI‘s measurements of the hard X-ray continuum are the best ever obtained for an astrophysical source. They are precise enough for model-independent deconvolution using powerful mathematical techniques developed for RHESSI observations (e.g.,\cite{2011SSRv..159..301K}) to obtain the spectrum of the bremsstrahlung-emitting source electrons. For flare hard X-ray spectra with a flattening at low energies, the derived electron source spectra appear to show a roll-off around 20~--40~keV \cite{2005SoPh..232...63KKasparovaKarlickyKontarSoPh2005}, but after correcting for albedo (the Compton scattering of the source hard X-rays by the solar photosphere), using a newly-developed Green’s function method \cite{2006A&A...446.1157KKontarMacKinnonSchwartzA&A2006}, all the derived source electron power-law spectra extend with no roll-off down to $\leq$20 keV and sometimes as low as $\sim$12~keV (see also \cite{2004ApJ...613.1233MMassoneEmslieKontarApJ2004,2005ApJ...626.1102SSuiHolmanDennisApJ2005,2007ApJ...670..862SSuiHolmanDennisApJ2007}), where the hot flare thermal emission dominates. These spectral measurements have also been used to demonstrate the so-called Neupert Effect, in which the time history of the thermal soft X-ray emission in many flares closely matches the time integral of the non-thermal hard X-ray emission \cite{2005ApJ...621..482VVeronigBrownDennisApJ2005}. 

Power-law spectra extending to such low energies imply that the source energetic electrons must contain a large fraction of the energy released in many flares \cite{2005A&A...435..743SSaint-HilaireBenzA&A2005}. Detailed energy budgets of solar eruptive events (SEEs) including the flare and associated coronal mass ejection have supported this result, e.g.  \cite{2012ApJ...759...71EEmslieDennisShihApJ2012} and a series of papers culminating in \cite{2017ApJ...836...17AAschwandenCaspiCohenApJ2017,2020ApJ...903...23AAschwandenApJ2020}. A critical comparison of the different published results by \cite{2020A&A...644A.172WWarmuthMannA&A2020} is summarised in Fig.~\ref{WarmuthMann}. While it is clear that the non-thermal energy in injected electrons is strongly dependent on the poorly constrained low-energy cutoff, consistent results for all flares in the study were obtained using a spectral cross-over method. However, the non-thermal energies for many flares reported by \cite{2017ApJ...836...17AAschwandenCaspiCohenApJ2017} clearly violate the limits imposed by the measured bolometric radiated energy that must be an upper limit on the total energy released in a flare.

\subsection{Non-thermal Emissions from the Corona, and Bulk Energisation}

\begin{figure}[pht]
\begin{center}
\includegraphics[width=1.0\textwidth, 
    angle = 0, trim = 0 0 0 0]
    {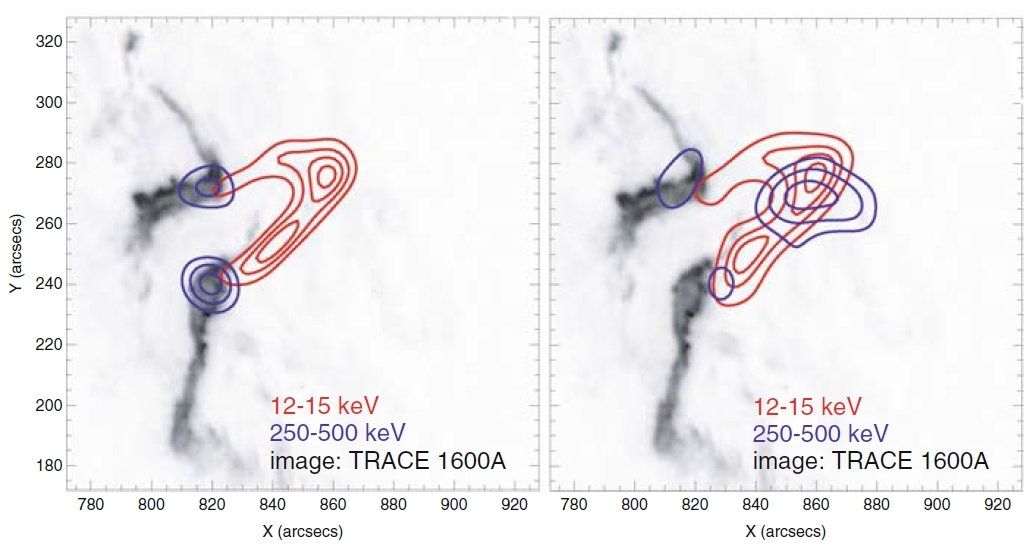}
\end{center}
\caption{X-ray images of flare on 20 January 2005 showing a coronal hard X-ray source at energies up to the 250--500 keV energy bin. From \cite{2008ApJ...678L..63KKruckerHurfordMacKinnonApJL2008}
}
\label{Krucker2008}
\end{figure}

RHESSI's excellent energy resolution, in combination with its imaging capability, made it possible for the first time to cleanly measure non-thermal emissions from the corona (e.g., \cite{2007A&A...466..713BBattagliaBenzA&A2007} and Fig.~\ref{Krucker2008}). These observations reveal that a non-thermal component is present in essentially all flares \cite{2008ApJ...673.1181KKruckerLinApJ2008}. RHESSI discovered that coronal non-thermal emissions are found in a large variety of associations such as in the pre-flare phase \cite{2003ApJ...595L..69LLinKruckerHurfordApJL2003}, in the absence of footpoint emission \cite{2004ApJ...603L.117VVeronigBrownApJL2004}, associated with jets \cite{2009A&A...508.1443BBainFletcherA&A2009}, with coronal mass ejections \cite{2007ApJ...669L..49KKruckerWhiteLinApJL2007}, and also in the gamma-ray range \cite{2008ApJ...678L..63KKruckerHurfordMacKinnonApJL2008}. These observations indicate that a significant fraction of the total accelerated electrons are in the corona at one time, clearly favouring an acceleration site in the corona.  Some rare, extremely bright sources provide crucial tests of the maximal efficiency of different acceleration models. The extreme brightness suggests that all electrons within the source are accelerated in a bulk energisation process \cite{2010ApJ...714.1108KKruckerHudsonGlesenerApJ2010, 2011ApJ...737...48IIshikawaKruckerTakahashiApJ2011}. Simultaneous microwave observations indicate that the energy in accelerated electrons at the peak of the event is of the same order of magnitude as the magnetic energy (i.e., plasma beta near unity). This indicates an extremely efficient conversion of magnetic energy into kinetic energy. 

\subsection{Double Coronal X-ray Sources}

\begin{figure}[pht]
\begin{center}
\includegraphics[width=1.0\textwidth, 
    angle = 0, trim = 0 0 0 0]
    {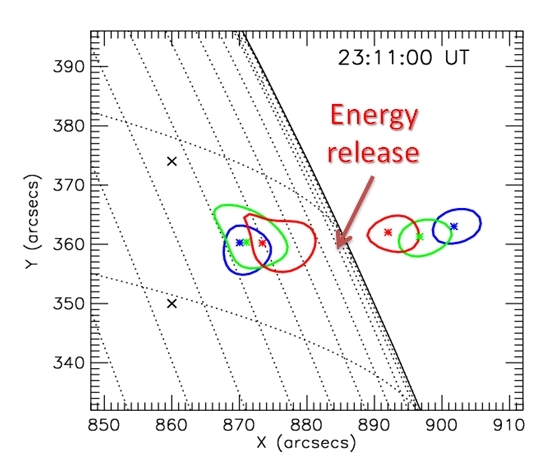}
\end{center}
\caption{RHESSI images made for a flare on 15 April 2002 at 23:11:00 UT showing twin thermal X-ray sources and the presumptive location of the coronal energy release site. Colour-coded contours at 80\% of the peak flux in each case, and asterisks marking the centroids, are shown for three energy bins.
For the contours to the left on the solar disk, the energy bins were 6–8, 10–12, and 16–20 keV. In this case, the source is presumed to be at the top of a large, symmetric magnetic loop joining the two footpoints seen in other images and marked here with crosses. For the contours to the right above the solar limb, the energy bins were 10--12, 12--14, and 14--16 keV. In each case the colour coding was blue, green, and red, respectively. From \cite{2003ApJ...596L.251SSuiHolmanApJL2003} with permission of L. Sui.
}
\label{SuiHolman2003}
\end{figure}

The detection of twin coronal X-ray sources, one at the top and the other above the top of the flare loops, was first reported by \cite{2003ApJ...596L.251SSuiHolmanApJL2003}. As shown in Fig.~\ref{SuiHolman2003}, for both coronal sources, the location of higher energy, and therefore higher temperature, emission is shifted toward the region between the sources.  This provided strong evidence for energy release in the corona between the two sources. Several other similar events have been reported (e.g., \cite{2008ApJ...676..704LLiuPetrosianDennisApJ2008}), but it is not known how common this phenomenon might be.

\subsection{Initial Downward Motion of X-ray Sources}

\begin{figure}[pht]
\begin{center}
\includegraphics[width=1.0\textwidth, 
    angle = 0, trim = 0 0 0 0]
    {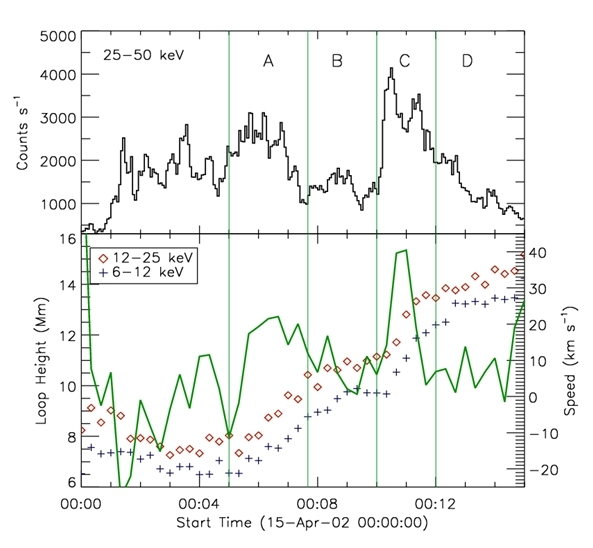}
\end{center}
\caption{Altitude variations of the coronal X-ray source observed during the 14–15 April 2002 flare. From \cite{2004ApJ...612..546SSuiHolmanDennisApJ2004}
}
\label{SuiHolmanDennis2004}
\end{figure}

Sui and Holman \cite{2003ApJ...589L..55S} reported the initial downward motion of the higher altitude coronal X-ray source seen in Fig.~\ref{SuiHolman2003} prior to the previously reported continuous upward motion. This initial downward motion was further detailed by \cite{2004ApJ...612..546SSuiHolmanDennisApJ2004} as shown in Fig.~\ref{SuiHolmanDennis2004}. This figure also shows that the rate of altitude increase correlates with the hard X-ray flux, suggesting that it was related to the energy release rate. The initial downward motion has been seen now in other flares (e.g., \cite{2004ApJ...612..546SSuiHolmanDennisApJ2004,2006A&A...446..675VVeronigKarlickyVrsnakA&A2006,2008ApJ...680..734JJiWangLiuApJ2008}) and at other wavelengths, and it appears to be associated with the propagation of reconnection along flare ribbons, but its interpretation is still unclear.

\subsection{Microflares and the Quiet Sun}

\begin{figure}[pht]
\begin{center}
\includegraphics[width=1.0\textwidth, 
    angle = 0, trim = 0 0 0 0]
    {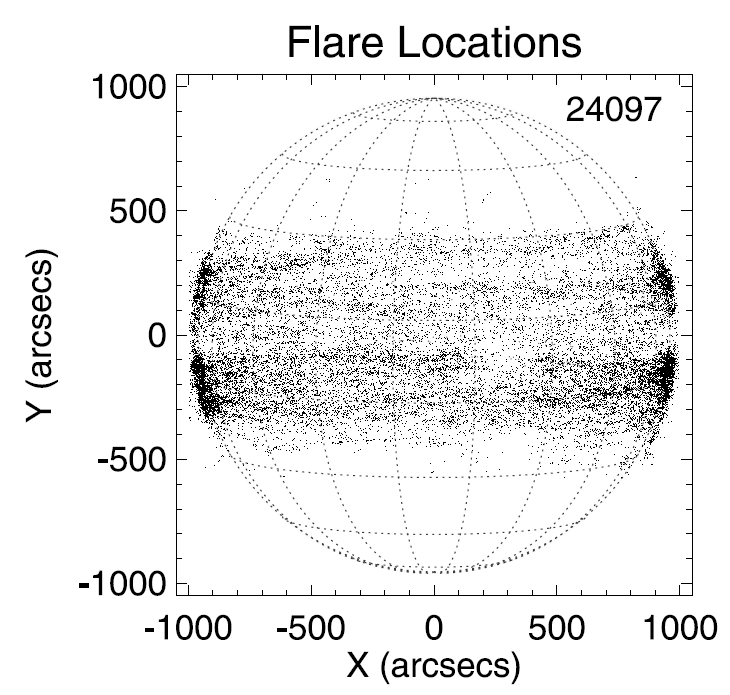}
\end{center}
\caption{Location on the solar disk of $\sim$25,000 \added{transient X-ray} microflares detected with RHESSI in the 6--12~keV energy band.  From \cite{2008ApJ...677.1385CChristeHannahKruckerApJ2008} with permission of S. Christe.}
\label{Christeetal2008}
\end{figure}

RHESSI could readily observe microflares in both hard and soft X-rays,
determining both their spectrum and location on the solar disk \cite{2002SoPh..210..445KKruckerChristeLinSoPh2002,2002SoPh..210..431BBenzGrigisSoPh2002,2007SoPh..246..339SStoiserVeronigAurassSoPh2007,2008ApJ...677.1385CChristeHannahKruckerApJ2008, 2008ApJ...677..704HHannahChristeKruckerApJ2008,2011SSRv..159..263HHannahHudsonBattagliaSSRv2011}.
Figure~\ref{Christeetal2008} shows the locations of some 25,000 transient X-ray microflares detected with RHESSI in the 6--12~keV energy band. Their distribution in space shows that they all come from active regions within the active North and South latitude bands. This establishes conclusively that transient X-ray microflares such as these do not heat the quiet corona. However, an important caveat to this result is that the  detection methodology used by \cite{2008ApJ...677.1385CChristeHannahKruckerApJ2008} fails to detect microflares that are not sufficiently transient.

Related to this, RHESSI has established the lowest upper limits on quiet-Sun hard X-ray emission in the absence of active regions \cite{2010ApJ...724..487HHannahHudsonHurfordApJ2010}, placing strong constraints on any non-thermal energy release by nanoflares or other phenomena, and even on axions produced in the core of the Sun \cite{2012ASPC..455...25HHudsonActonDeLucaASPC2012}.

\subsection{Timing of HXR Flare Ribbons}

\begin{figure}[pht]
\begin{center}
\includegraphics[width=1.0\textwidth, 
    angle = 0, trim = 0 0 0 0]
    {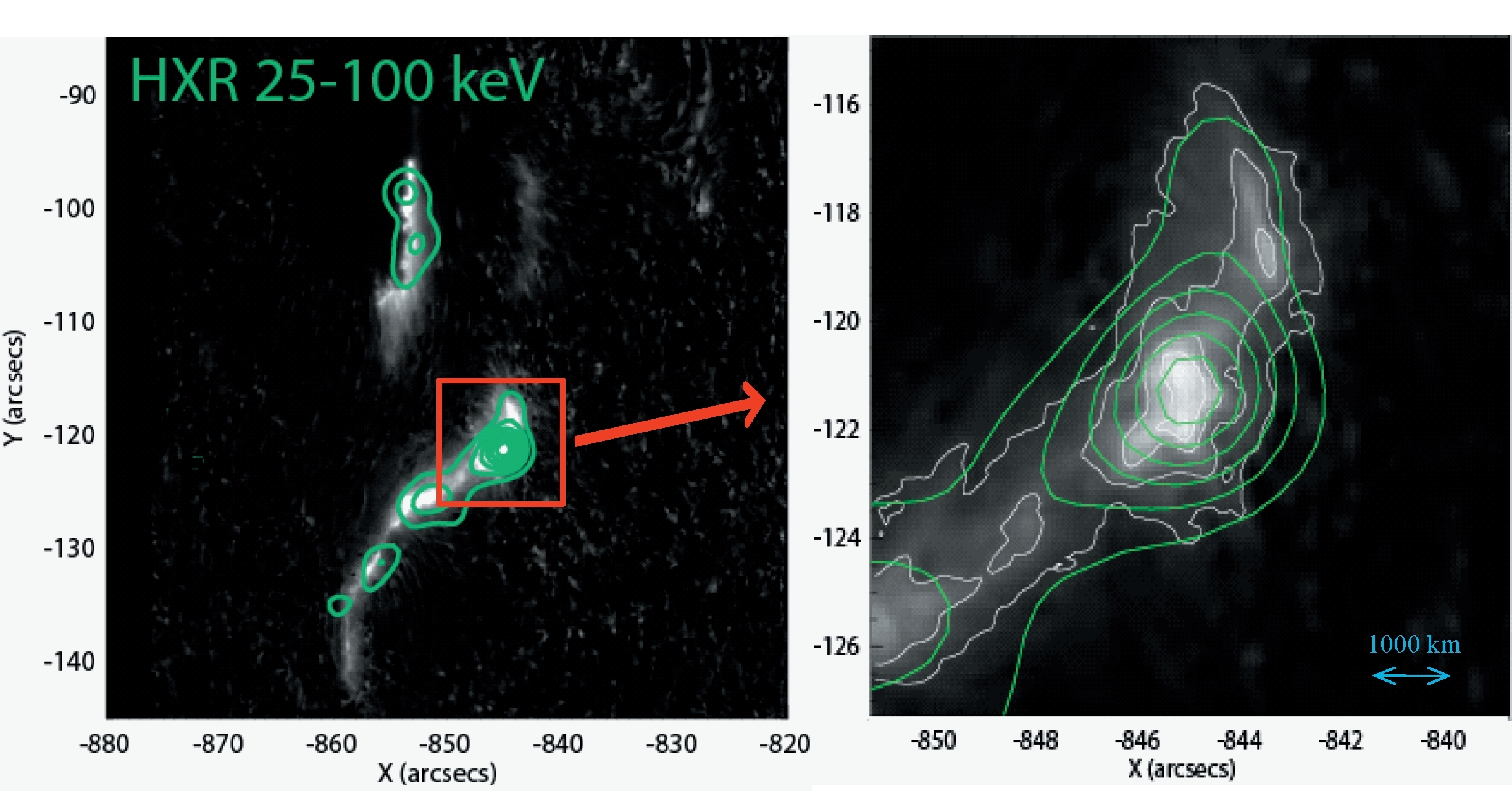}
\end{center}
\caption{Hard X-ray (green contours) and white light (G-band) imaging of a flare ribbon for a flare on 2006 December 6 reported by \cite{2011ApJ...739...96KKruckerHudsonJeffreyApJ2011}. The left image shows the two ribbons. The right image is an enlargement of the brightest part of the southern ribbon. From RHESSI Science Nugget \#142 available from \url{https://hesperia.gsfc.nasa.gov/rhessi3/news-and-resources/results/nuggets/}.
}
\label{HXRFlareRibbons}
\end{figure}

With RHESSI’s unprecedented ($\sim$2 arcsec) spatial resolution, detailed hard X-ray imaging of flare ribbons finally became possible \cite{2002SoPh..210..307FFletcherHudsonSoPh2002}\footnote{See also RHESSI Science Nugget \#142 available from \url{https://hesperia.gsfc.nasa.gov/rhessi3/news-and-resources/results/nuggets/}}. An example is shown in Fig.~\ref{HXRFlareRibbons}. Elongated HXR structures along the ribbons closely match the white light sources but only partially match the EUV sources (\cite{2007ApJ...658L.127LLiuLeeGaryApJL2007,2008ApJ...676..704LLiuPetrosianDennisApJ2008,2009ApJ...698.2131DDennisPernakApJ2009,2011ApJ...739...96KKruckerHudsonJeffreyApJ2011}). 
The width of the hard X-ray sources, however, was often unresolved even by RHESSI, yielding an upper limit of $\sim$1 arcsec for the dimension \cite{2009ApJ...698.2131DDennisPernakApJ2009}. This indicates that the number density of precipitating electrons is even larger than previously thought. These observations challenge our current understanding of the standard thick-target model and contribute new insights to our understanding of electron acceleration and transport. 

\subsection{Location of Super-hot X-ray Sources}

%\todo{This subsection should probably be updated with the analysis of Caspi et al. 2015, which produced separate images of the two thermal components}

\begin{figure}[pht]
\begin{center}
\includegraphics[width=0.8\textwidth, 
    angle = 0, trim = 0 0 0 0]
    {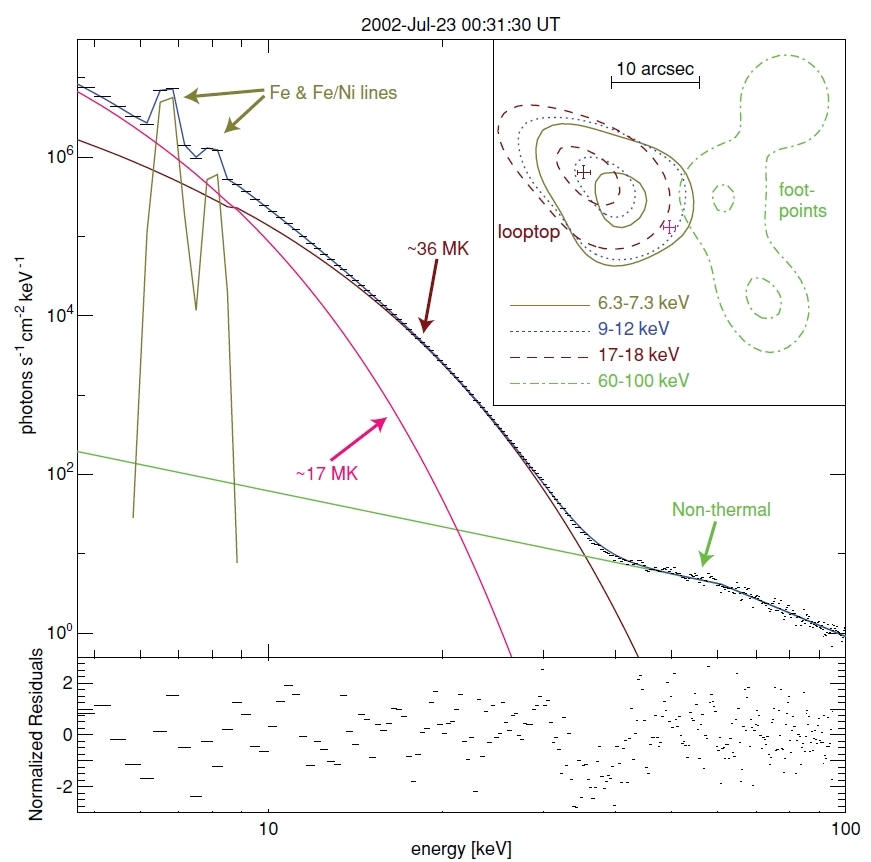}
\end{center}
\caption{X-ray spectrum and image of the 2002 July 23 X4.8 flare. From \cite{2010ApJ...725L.161CCaspiLinApJL2010} with permission of A. Caspi.
}
\label{CaspiLin}
\end{figure}

Super-hot flare plasmas, defined as plasma with temperatures in excess of 30~MK, were discovered about three decades ago.  RHESSI’s high spectral and spatial resolution has allowed the soft X-ray emission during the 2002 July 23 X4.8 flare to be separated into two spatially distinct isothermal coronal X-ray sources as shown in Fig.~\ref{CaspiLin}. A super-hot source with a temperature of $\sim$36 MK is located high in the corona, with a cooler $\sim$17 MK source located at a lower altitude \cite{2010ApJ...725L.161CCaspiLinApJL2010}. The analysis was improved by \cite{2015ApJ...811L...1CCaspiShihMcTiernanApJL2015} using a new analytical technique that combined visibilities at different energies to allow images of the solar flare emission to be displayed as a function of the isothermal temperature rather than photon energy. The super-hot source was present at high altitude even during the flare pre-impulsive phase when no HXR footpoint emission was detected. It was farther from the footpoints and more elongated throughout the impulsive phase, consistent with an in~situ heating mechanism for the super-hot plasma while the ``normal'' $\sim$20 MK plasma originated primarily from chromospheric evaporation. The super-hot and hot plasmas thus arise from fundamentally different physical processes.

\subsection{The Photosphere as a Compton or "Dentist's" Mirror}

%\begin{figure}[pht]
%\begin{center}
%\includegraphics[width=0.8\textwidth, 
%    angle = 0, trim = 0 0 0 0]
%    {figures/KontarBrown2006.jpg}
%\end{center}
%caption{Spectra of upwards and downwards directed electrons.}
%\label{KontarBrown}
%\end{figure}

With RHESSI's excellent spectral resolution, we can use the albedo contribution to the measured hard X-ray flux to our advantage.  \cite{2006ApJ...648.1239K} showed that, by considering the Sun’s surface to act as a “Compton mirror,” we can look at the emitting electrons both directly and from behind the source, providing vital information on the directionality of the propagating particles\footnote{See RHESSI Science Nugget \#42 available from \url{https://hesperia.gsfc.nasa.gov/rhessi3/news-and-resources/results/nuggets/}}. Using this technique, they determined simultaneously the electron spectra of both the downward- and the upward-directed electrons for two solar flares observed with RHESSI. The results 
%shown in Fig.~\ref{KontarBrown} 
reveal surprisingly near-isotropic electron distributions, which contrast strongly with the expectations from the standard model that invokes strong downward beaming, including a collisional thick-target model. This result suggest that simple electron beams as normally envisioned aren't present in the energetically important distributions.

\subsection{Broadened 511-keV Positron Annihilation Line}

\begin{figure}[pht]
\begin{center}
\includegraphics[width=0.8\textwidth, 
    angle = 0, trim = 0 0 0 0]
    {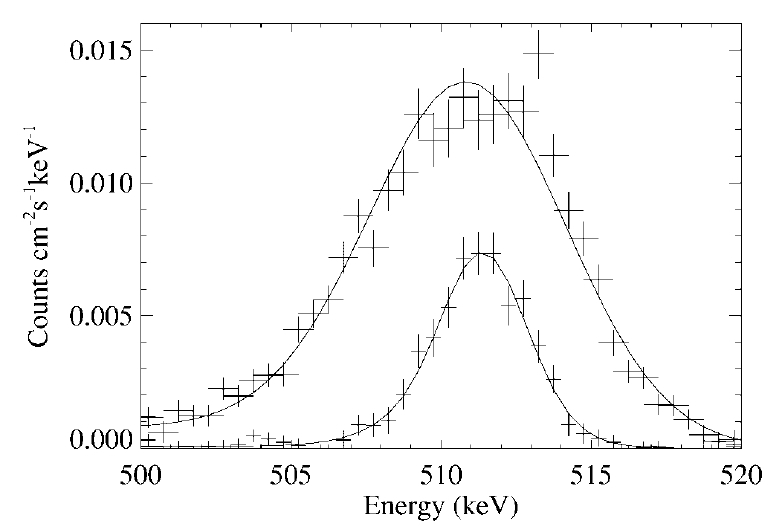}
\end{center}
\caption{RHESSI count spectra of the solar 511 keV annihilation line at two times during the 28 October 2003 flare. From \cite{2004ApJ...615L.169SShareMurphySmithApJL2004} with permission of G.~Share.
}
\label{Shareetal}
\end{figure}

RHESSI’s high resolution spectroscopy of the flare 511-keV positron annihilation line emission showed a line width of typically $\ge$5 keV, indicating that the temperature of the accelerated-ion interaction region was above $10^5$ K. Later, in some flares, as shown in Fig.~\ref{Shareetal}, the width of the line narrows to $\sim$1 keV, consistent with annihilation in ionised hydrogen at $<10^4$K and $\ge10^{15} cm^{-3}$. 
The full implications of these observations are still unclear but they bring into question the energy source of the heating. We also do not know whether or how the ions alone can produce such a highly dynamic flaring atmosphere at chromospheric densities that can reach transition-region temperatures with large column depths, and then cool to less than $10^4$~K in minutes while remaining highly ionised.

%\subsection{Non-flare-related Results}

\subsection{Solar Oblateness}

\begin{figure}[pht]
\begin{center}
\includegraphics[width=0.8\textwidth, 
    angle = 0, trim = 0 0 0 0]
    {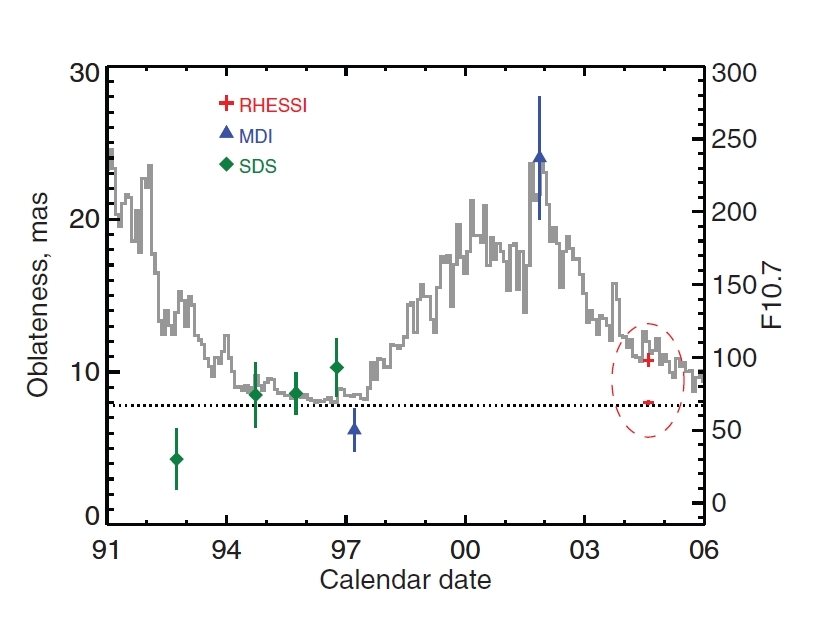}
\end{center}
\caption{Comparison of oblateness measurements from space. The RHESSI oblateness measurements are from data from 29 June to 24 September 2004 (red crosses). Results are also shown from the balloon-borne Solar Disk Sextant (SDS) (green diamonds) and the MDI instrument on board SOHO (blue triangles). The surface rotation rate predicts the value shown with the dotted line. The histogram (scaled to the uniform-rotation oblateness at solar minimum and to the higher MDI data point) shows the radio flux index F10.7, a good indicator of the solar cycle. From \cite{2008Sci...322..560FFivianHudsonLinSci2008} with permission of H.~Hudson.
}
\label{Fivianetal}
\end{figure}

\cite{2008Sci...322..560FFivianHudsonLinSci2008} used data from RHESSI's Solar Aspect System to provide the most precise measurement of the shape of the Sun\footnote{See RHESSI Science Nugget \#126 available from \url{https://hesperia.gsfc.nasa.gov/rhessi3/news-and-resources/results/nuggets/}.} shown in Fig.~\ref{Fivianetal}. Initially, the raw measurements showed an unexpectedly large flattening compared to what is predicted from solar rotation. However, this effect was likely due to magnetic elements in the enhanced network producing emission seen at the solar limbs, preferentially at lower latitudes. Once this contribution was removed, the corrected oblateness of the non-magnetic Sun was determined to be $8.01 \pm 0.14$ milli–arc seconds, which is near the value of 7.8 milli–arc seconds, or $\sim$0.001\%, expected from solar rotation. This result is the most accurate measure of the true oblateness ever made and may explain the apparent variation with the solar cycle that had been reported earlier.

\subsection{Magnetar Timing and Spectroscopy}

\begin{figure}[pht]
\begin{center}
\includegraphics[width=0.8\textwidth, 
    angle = 0, trim = 0 0 0 0]
    {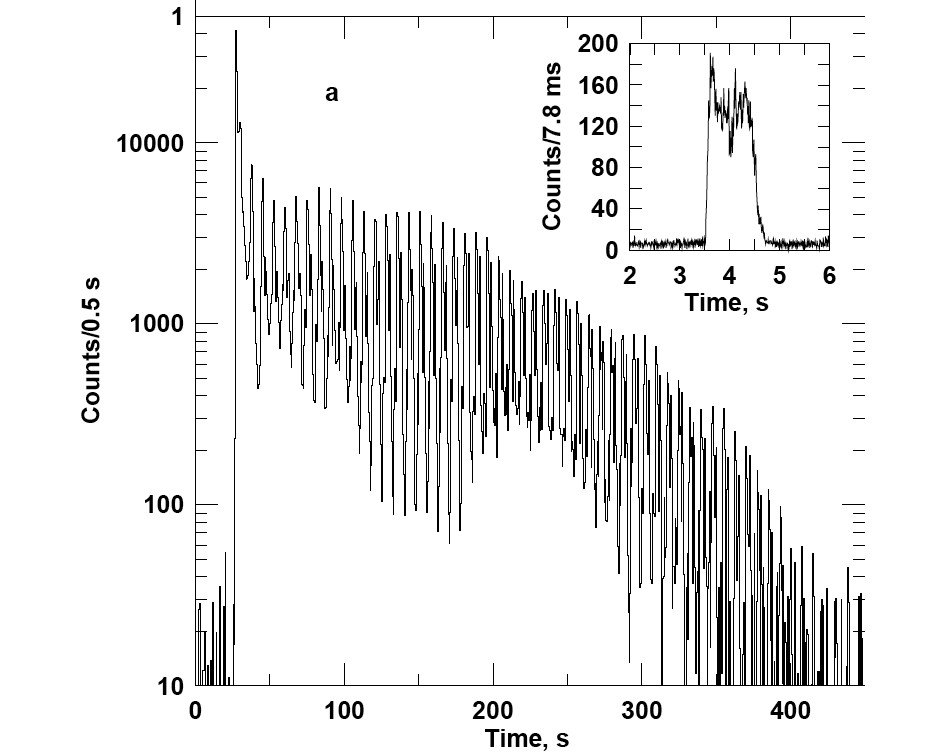}
\end{center}
\caption{RHESSI 20--100 keV light curves of the giant magnetar flare on 27 December 2004 plotted with 0.5-s time resolution. From \cite{2005Natur.434.1098HHurleyBoggsSmithNatur2005}}
\label{Magnetar}
\end{figure}

RHESSI serendipitously detected a huge flare on 27 December 2004\footnote{See RHESSI Science Nugget \#3 available from \url{https://hesperia.gsfc.nasa.gov/rhessi3/news-and-resources/results/nuggets/}} with the detailed light curve shown in Fig.~\ref{Magnetar}. It was determined to have come from the soft-gamma-ray repeater (SGR) 1806–20 that was just 5.25$\degree$ from the Sun at that time \cite{2005Natur.434.1098HHurleyBoggsSmithNatur2005}.
 
SGRs are thought to be magnetars - isolated, strongly magnetised neutron stars with tera-Gauss exterior magnetic fields and even stronger interior fields, making them the most strongly-magnetised objects in the Universe. In the first 0.2 s of the event, the flare released as much energy as the Sun radiates in a quarter of a million years. This observation suggested that a significant fraction of the mysterious short-duration gamma-ray bursts may come from similar extragalactic magnetars.

Later work by \cite{2006ApJ...637L.117WWattsStrohmayerApJL2006} revealed quasi-periodic oscillations (QPOs) in the RHESSI timing data, directly related to the "ringing" modes of the neutron star. When interpreted as arising from vibrations in the neutron star crust, these QPOs offer a novel means of testing the neutron star equation of state, crustal breaking strain, and magnetic field configuration.

\subsection{Terrestrial Gamma-ray Flashes (TGFs)}

\begin{figure}[pht]
\begin{center}
\includegraphics[width=0.8\textwidth, 
    angle = 0, trim = 0 0 0 0]
    {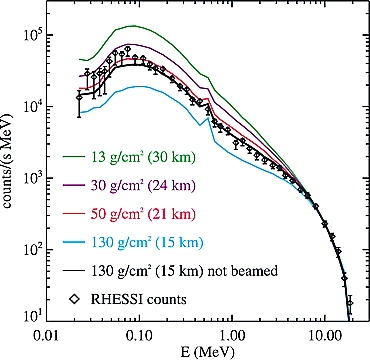}
\end{center}
\caption{TGF counts spectrum as measured by RHESSI and the X-ray emission spectra, corrected for the instrumental response, as calculated by the Monte Carlo simulation of runaway breakdown for E/n = 400 kV/m at four atmospheric depths. From \cite{2005GeoRL..3222804DDwyerSmithGeoRL2005} with permission of D. Smith.}
\label{DwyerSmith}
\end{figure}

TGFs were first discovered with BATSE on the Compton Gamma Ray Observatory and were associated with terrestrial lightning. RHESSI has detected many TGFs \cite{2005Sci...307.1085SSmithLopezLinSci2005} showing that they are much more common and luminous than previously thought.\footnote{RHESSI Science Nugget \#32 available from \url{https://hesperia.gsfc.nasa.gov/rhessi3/news-and-resources/results/nuggets/}} They extend up to gamma-ray energies beyond 20 MeV as shown in Fig.~\ref{DwyerSmith}, with a spectral shape predicted by the relativistic runaway model \cite{2005GeoRL..3222804DDwyerSmithGeoRL2005}. RHESSI's observations also show that TGFs come from altitudes of around 15 km, and that they are associated with intra-cloud lightning and not cloud-to-ground lightning or sprites.\\\\\

%\begin{Acknowledgments}

\noindent\textbf{Acknowledgments}\\

We acknowledge four individuals who contributed so much to RHESSI's tremendous scientific success but who are sadly no longer with us: Reuven Ramaty, a pioneer in the field of gamma-ray astronomy (the “R” in RHESSI honors his contributions); Bob Lin, the original RHESSI Principal Investigator, whose scientific vision and tenacious energy was the major factor in the initiation and implementation of the unparalleled RHESSI mission; John Brown, whose pioneering theoretical studies into the relationship between solar hard X-rays and the electrons that produce them set the stage for the RHESSI mission; and Richard Schwartz, whose unique combination of technical knowledge and scientific insight was responsible for much of the RHESSI data analysis software and computational algorithms that have been used by mere mortals trying to understand the observations in their scientific context. We thank Kim Tolbert for proofreading the manuscript and for continuing to improve the RHESSI documentation and data analysis software and for making it so user-friendly. Any inaccuracies, or errors of omission or commission, that remain in this paper are, of course, the responsibility of the authors.
%\end{Acknowledgments}

%\begin{thebibliography}{99.}
\bibliography{references,BRD_references,HXR_Imaging}
%\end{thebibliography}

\end{document}